\documentclass[aps,prd,altaffilletter,reprint,amsmath,amssymb,
superscriptaddress,showpacs]{revtex4-1}
\usepackage{graphicx}
\usepackage{dcolumn}
\usepackage{bm}
\usepackage{bbding}
\usepackage{multirow}
\usepackage{color}
\begin{document}

\title{Search for heavy neutrinos in $K^+\to\mu^+\nu_H$ decays}

\author{A.V.~Artamonov}\affiliation{Institute for High Energy Physics,
Protvino,
Moscow Region, 142 280, Russia}
\author{B.~Bassalleck}\affiliation{Department of Physics and Astronomy,
University of New Mexico, Albuquerque, NM 87131}
\author{B.~Bhuyan}
\altaffiliation{Now at Department of Physics, Indian Institute of
Technology Guwahati, Guwahati, Assam, 781 039, India.}
\affiliation{Brookhaven
National Laboratory, Upton, NY 11973}
\author{E.W.~Blackmore}\affiliation{TRIUMF, 4004 Wesbrook Mall, Vancouver,
British Columbia, Canada V6T 2A3}
\author{D.A.~Bryman} \affiliation{Department of Physics and Astronomy,
University of British Columbia, Vancouver, British Columbia, Canada V6T 1Z1}
\author{S.~Chen} \affiliation{Department of Engineering Physics, Tsinghua
University, Beijing 100084, China} \affiliation{TRIUMF, 4004 Wesbrook Mall,
Vancouver, British Columbia, Canada V6T 2A3}
\author{I-H.~Chiang} \affiliation{Brookhaven National Laboratory, Upton, NY
11973}
\author{I.-A.~Christidi}
\altaffiliation{Now at Physics Department, Aristotle
University of Thessaloniki, Thessaloniki 54124, Greece.}
\affiliation{Department of
Physics and Astronomy, Stony Brook University, Stony Brook, NY 11794}
\author{P.S.~Cooper}\affiliation{Fermi National Accelerator Laboratory,
Batavia, IL 60510}
\author{M.V.~Diwan} \affiliation{Brookhaven National Laboratory, Upton, NY
11973}
\author{J.S.~Frank}
\altaffiliation{Now at 1 Nathan Hale Drive, Setauket, New York 11733.}
\affiliation{Brookhaven National Laboratory, Upton, NY 11973}
\author{T.~Fujiwara}\affiliation{Department of Physics, Kyoto University,
Sakyo-ku, Kyoto 606-8502, Japan}
\author{J.~Hu} \affiliation{TRIUMF, 4004 Wesbrook Mall, Vancouver, British
Columbia, Canada V6T 2A3}
\author{J.~Ives} \affiliation{Department of Physics and Astronomy, University
of British Columbia, Vancouver, British Columbia, Canada V6T 1Z1}
\author{A.O.~Izmaylov}\affiliation{Institute for Nuclear Research RAS, 60
October Revolution Prospect 7a, 117312 Moscow, Russia}
\author{D.E.~Jaffe} \affiliation{Brookhaven National Laboratory, Upton, NY
11973}
\author{S.~Kabe}
\altaffiliation{Deceased.}
\affiliation{High Energy Accelerator Research Organization~(KEK), Oho, Tsukuba,
Ibaraki 305-0801, Japan}
\author{S.H.~Kettell} \affiliation{Brookhaven National Laboratory, Upton, NY
11973}
\author{M.M.~Khabibullin}\affiliation{Institute for Nuclear Research RAS, 60
October Revolution Prospect 7a, 117312 Moscow, Russia}
\author{A.N.~Khotjantsev}\affiliation{Institute for Nuclear Research RAS, 60
October Revolution Prospect 7a, 117312 Moscow, Russia}
\author{P.~Kitching} \affiliation{Centre for Subatomic Research, University of
Alberta, Edmonton, Canada T6G 2N5}
\author{M.~Kobayashi}
\affiliation{High Energy Accelerator Research Organization~(KEK), Oho, Tsukuba,
Ibaraki 305-0801, Japan}
\author{T.K.~Komatsubara}
\affiliation{High Energy Accelerator Research Organization~(KEK), Oho, Tsukuba,
Ibaraki 305-0801, Japan}
\author{A.~Konaka} \affiliation{TRIUMF, 4004 Wesbrook Mall, Vancouver, British
Columbia, Canada V6T 2A3}
\author{Yu.G.~Kudenko}\affiliation{Institute for Nuclear Research RAS, 60
October Revolution Prospect 7a, 117312 Moscow, Russia}\affiliation{Moscow
Institute of
Physics and Technology, 141700 Moscow, Russia}
\affiliation{National Research Nuclear University MEPhI (Moscow Engineering
Physics Institute), 115409 Moscow, Russia}
\author{L.G.~Landsberg}\altaffiliation{Deceased.}\affiliation{Institute for
High Energy Physics, Protvino, Moscow Region, 142 280, Russia}
\author{B.~Lewis}\affiliation{Department of Physics and Astronomy, University
of New Mexico, Albuquerque, NM 87131}
\author{K.K.~Li}\affiliation{Brookhaven National Laboratory, Upton, NY 11973}
\author{L.S.~Littenberg} \affiliation{Brookhaven National Laboratory, Upton, NY
11973}
\author{J.A.~Macdonald} \altaffiliation{Deceased.} \affiliation{TRIUMF, 4004
Wesbrook Mall, Vancouver, British Columbia, Canada V6T 2A3}
\author{J.~Mildenberger} \affiliation{TRIUMF, 4004 Wesbrook Mall, Vancouver,
British Columbia, Canada V6T 2A3}
\author{O.V.~Mineev}\affiliation{Institute for Nuclear Research RAS, 60 October
Revolution Prospect 7a, 117312 Moscow, Russia}
\author{M. Miyajima} \affiliation{Department of Applied Physics, Fukui
University, 3-9-1 Bunkyo, Fukui, Fukui 910-8507, Japan}
\author{K.~Mizouchi}\affiliation{Department of Physics, Kyoto University,
Sakyo-ku, Kyoto 606-8502, Japan}
\author{N.~Muramatsu}\altaffiliation{Now at Research Center for Electron Photon
Science, Tohoku University, Taihaku-ku, Sendai, Miyagi 982-0826,
Japan.}\affiliation{Research Center for Nuclear Physics, Osaka University, 10-1
Mihogaoka, Ibaraki, Osaka 567-0047, Japan}
\author{T.~Nakano}\affiliation{Research Center for Nuclear Physics, Osaka
University, 10-1 Mihogaoka, Ibaraki, Osaka 567-0047, Japan}
\author{M.~Nomachi}\affiliation{Laboratory of Nuclear Studies, Osaka
University, 1-1 Machikaneyama, Toyonaka, Osaka 560-0043, Japan}
\author{T.~Nomura}\altaffiliation{Now at High Energy Accelerator Research
Organization (KEK), Oho, Tsukuba, Ibaraki 305-0801,
Japan.}\affiliation{Department of Physics, Kyoto University, Sakyo-ku,
Kyoto 606-8502, Japan}
\author{T.~Numao} \affiliation{TRIUMF, 4004 Wesbrook Mall, Vancouver, British
Columbia, Canada V6T 2A3}
\author{V.F.~Obraztsov}\affiliation{Institute for High Energy Physics,
Protvino, Moscow Region, 142 280, Russia}
\author{K.~Omata}
\affiliation{High Energy Accelerator Research Organization~(KEK), Oho, Tsukuba,
Ibaraki 305-0801, Japan}
\author{D.I.~Patalakha}\affiliation{Institute for High Energy Physics,
Protvino, Moscow Region, 142 280, Russia}
\author{R.~Poutissou} \affiliation{TRIUMF, 4004 Wesbrook Mall, Vancouver,
British Columbia, Canada V6T 2A3}
\author{G.~Redlinger} \affiliation{Brookhaven National Laboratory, Upton, NY
11973}
\author{T.~Sato}
\affiliation{High Energy Accelerator Research Organization~(KEK), Oho, Tsukuba,
Ibaraki 305-0801, Japan}
\author{T.~Sekiguchi}
\affiliation{High Energy Accelerator Research Organization~(KEK), Oho, Tsukuba,
Ibaraki 305-0801, Japan}
\author{A.T.~Shaikhiev}\affiliation{Institute for Nuclear Research RAS, 60
October Revolution Prospect 7a, 117312 Moscow, Russia}
\author{T.~Shinkawa} \affiliation{Department of Applied Physics, National
Defense Academy, Yokosuka, Kanagawa 239-8686, Japan}
\author{R.C.~Strand} \affiliation{Brookhaven National Laboratory, Upton, NY
11973}
\author{S.~Sugimoto} \altaffiliation{Deceased.}
\affiliation{High Energy Accelerator Research Organization~(KEK), Oho, Tsukuba,
Ibaraki 305-0801, Japan}
\author{Y.~Tamagawa} \affiliation{Department of Applied Physics, Fukui
University, 3-9-1 Bunkyo, Fukui, Fukui 910-8507, Japan}
\author{R.~Tschirhart} \affiliation{Fermi National Accelerator Laboratory,
Batavia, IL 60510}
\author{T.~Tsunemi} \altaffiliation{Now at Department of Physics, Kyoto
University, Sakyo-ku, Kyoto 606-8502, Japan.}
\affiliation{High Energy Accelerator Research Organization~(KEK), Oho, Tsukuba,
Ibaraki 305-0801, Japan}
\author{D.V.~Vavilov} \altaffiliation{Now at TRIUMF, 4004 Wesbrook Mall,
Vancouver, British Columbia, Canada V6T 2A3.}\affiliation{Institute for High
Energy Physics, Protvino, Moscow Region, 142 280, Russia}
\author{B.~Viren} \affiliation{Brookhaven National Laboratory, Upton, NY 11973}
\author{Zhe~Wang} \affiliation{Department of Engineering Physics, Tsinghua
University, Beijing 100084, China} \affiliation{Brookhaven National Laboratory,
Upton, NY 11973}
\author{Hanyu~Wei} \affiliation{Department of Engineering Physics, Tsinghua
University, Beijing 100084, China}
\author{N.V.~Yershov}\affiliation{Institute for Nuclear Research RAS, 60
October Revolution Prospect 7a, 117312 Moscow, Russia}
\author{Y.~Yoshimura}
\affiliation{High Energy Accelerator Research Organization~(KEK), Oho, Tsukuba,
Ibaraki 305-0801, Japan}
\author{T.~Yoshioka}
\altaffiliation{Now at  Department of Physics, Kyushu University, Higashi-ku,
Fukuoka 812-8581, Japan.}
\affiliation{High Energy Accelerator Research Organization~(KEK), Oho, Tsukuba,
Ibaraki 305-0801, Japan}
\collaboration{E949 Collaboration}\noaffiliation

\date{\today}

\begin{abstract}

Evidence of a heavy neutrino, $\nu_H$, in the $K^+\to\mu^+\nu_H$ decays was
sought using the E949 experimental data with an exposure of $1.70\times
10^{12}$ stopped kaons.
With the major background from the radiative $K^+\to\mu^+\nu_{\mu}\gamma$
decay understood and suppressed, upper limits (90\% C.L.) on the neutrino
mixing matrix
element between muon and heavy neutrino, $|U_{\mu H}|^2$, were
set at the level of $10^{-7}$ to $10^{-9}$
for the heavy neutrino mass region 175 to 300~MeV/$c^2$.

\end{abstract}

\pacs{14.60.St, 13.20.Eb}
\keywords{heavy neutrino, kaon decay, E949}

\maketitle

\section{Introduction \label{sec:intro}}
With neutrino mass and mixing
confirmed~(see~\cite{ref1,ref2,ref3,ref4,ref5,ref6,ref7,ref8,ref9,ref10,ref11,
ref12,ref13,ref14,ref15,ref16}~and references therein), a natural extension of
the Standard Model~(SM) involves the inclusion of sterile neutrinos which mix
with
ordinary neutrinos to explain phenomena that may be inconsistent with the
Standard Model.
An example of such a theory is the Neutrino Minimal Standard Model
($\nu$MSM)~\cite{nuMSM1,nuMSM2}. In this model, three massive right-handed
neutrinos are introduced to explain simultaneously neutrino oscillations, dark matter
and baryon asymmetry of the Universe.

The weak eigenstates of the neutrinos are related to their mass eigenstates by
a unitary matrix. The active neutrino mixing  can be  induced through

\begin{equation}
    \nu_l = \sum_{i} U_{li}\nu_i + \sum_{I}(U_{lH})_I(\nu_H)_I,
\end{equation}
where $l = e,\mu,\tau $; the mass eigenstates with masses $m_i$ are denoted as
$\nu_i~(i=1,2,3)$; $U_{li}$ are the elements of the
Pontecorvo-Maki-Nakagawa-Sakata  matrix;  $(\nu_H)_I~(I=1,2,3)$ are heavy
sterile neutrinos with masses $M_I$;  and $(U_{lH})_I$  are mixing parameters
between active neutrinos and heavy neutrinos.

\begin{figure*}[]
\centering
\includegraphics[width=15cm]{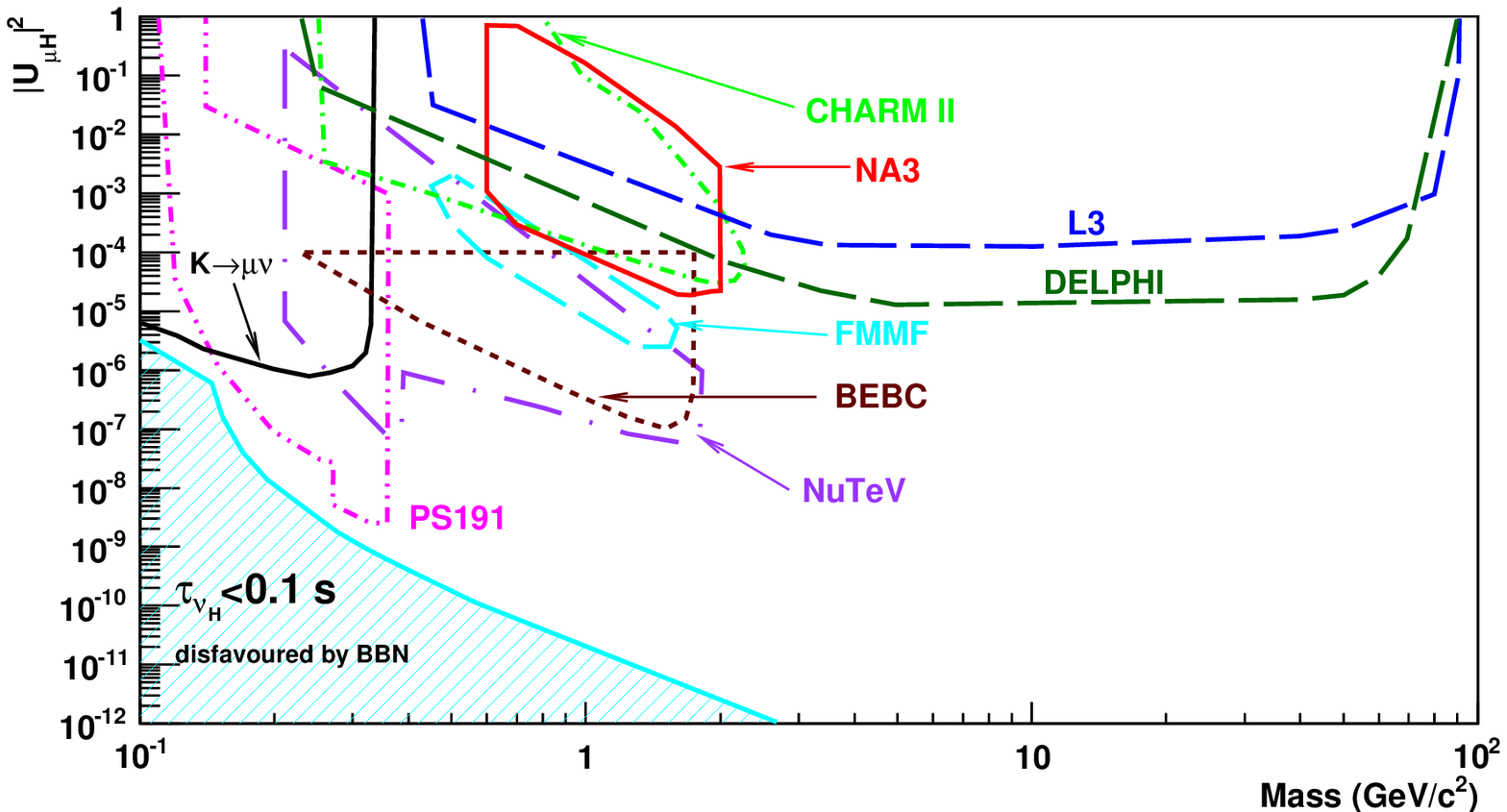}
\caption{(color online) Limits on $|U_{\mu H}|^2$ versus heavy neutrino mass in
the mass range
100 MeV/c$^2$--100 GeV/c$^2$. The
area with the solid (black) contour labeled $K\to\mu\nu$ is excluded by
production
searches~\cite{ex1}. The bounds by decay searches indicated by contours labeled
by
PS191~\cite{ps191}, NA3~\cite{na3}, BEBC~\cite{bebc}, FMMF~\cite{fmmf},
NuTeV~\cite{nutev} and CHARMII~\cite{charm2} are at 90~\%~C.L., while
DELPHI~\cite{delphi} and L3~\cite{l3} are at 95~\%~C.L. and are deduced from
searches of visible products in heavy neutrino decays. The shaded region shows
one of the possible lower bounds from Big Bang
Nucleosynthesis~\cite{prediction, prediction2}.}
\label{fig:constraints}
\end{figure*}

In the $\nu$MSM the mixing between active light neutrinos and  heavy sterile
neutrinos   gives rise to the  production of heavy neutrinos in weak decays of
heavy mesons as well as the decay of the heavy neutrinos to SM particles. Two
strategies can be used in the experimental searches for these particles. The
first one is related to their production. Since they are massive, the
kinematics of two body decays $K^\pm\to\mu^\pm\nu_\mu$ ($K_{\mu 2}$) and
$K^\pm\to\mu^\pm\nu_H$ are not the same. The study of kinematics of rare 
meson
decays can constrain the strength of the coupling of heavy leptons using the
following expression~\cite{shrock}:

\begin{equation}
\Gamma(M^+\to l^+\nu_H)=\rho\Gamma(M^+\to l^+\nu_l)|U_{lH}|^2,
\label{eq:u}
\end{equation}
where $M=\pi,\textnormal{ } K;\textnormal{ } l=e,\textnormal{ }\mu;\textnormal{
}\rho$ is a kinematical factor which lies in the range from 1 to 4 for
$0<m_{\nu_H}<300$~MeV/$c^2$; and $\nu_H$ is one of the $\nu$MSM heavy 
neutrinos.
This strategy has been used in a number of experiments for the search of neutral
leptons (heavy neutrinos) in the past~\cite{ex0,hayano,ex1,ex2,ex3}, where the
spectra of electrons and muons originating in decays of pions and kaons have
been studied. The second strategy is to look for the decays of heavy neutrinos
to hadrons and leptons~\cite{ps191, na3, bebc, fmmf, nutev, delphi, l3,
charm2}.
The bounds on mixing matrix element
$|U_{\mu H}|^2$ are shown in Figure~\ref{fig:constraints}~\cite{figure1}.

The best constraints in the small mass region $m_{\nu_H}<450$~MeV/$c^2$ were
from the CERN PS191~\cite{ps191} experiment, giving roughly $|U_{\mu
H}|^2<10^{-9}$ in the
region 250$<m_{\nu_H}<450$~MeV/$c^2$. In the CERN PS191
analysis, the heavy neutrino production and decay rates were calculated
for usual Dirac neutrinos with the sole assumption of a Dirac mass matrix.
Production rates for the channels presented in Figure~\ref{fig:constraints} are
not modified if the mass eigenstates are of the Majorana type. Decay rates for
non-self charge conjugate channels are multiplied by a factor of two for
Majorana neutrinos. The limits given here should therefore be divided by the
square root of two if one considers massive Majorana neutrinos.

The successful predictions of the Big Bang Nucleosynthesis~(BBN) also allow
establishment of  a number of lower bounds on the couplings of neutral
leptons~\cite{prediction, prediction2}, which considerably limit the allowable 
window for
the couplings and masses, but these bounds are model-dependent. The existence
of heavy neutrinos should not spoil the BBN predictions, so the heavy neutrino
lifetime, $\tau_{\nu_H}$, should be less than 0.1~s.
The BBN bound shown in Figure~\ref{fig:constraints} was calculated in the
$\nu$MSM framework assuming that the coupling of $\nu_H$ to the
third generation of leptons is stronger  than to the others. In case of
maximum coupling of $\nu_H$ to the first~(second) generation of leptons
the BBN bound will be weaker~(stronger) as shown in~\cite{prediction}.

In this paper, we present the result of a search for heavy neutrinos in
$K^+\to\mu^+\nu_H$ decays
from the inclusive muon spectrum of $K^+\to\mu^++anything$ decays using 
the kaon
decay-at-rest data from the E949~\cite{e949} experiment.
Since the E949 experiment focused on measuring the branching ratio of the
rare kaon decay $K^+\to\pi^+\nu\bar{\nu}$, the principal trigger selection
criteria were designed to identify pions and reject muons.
In the present analysis, however, decay product muons must be identified; these
muons are present in the data set due to inefficiencies in the pion selection
criteria applied. In this analysis we used E949 data taken from March to June
in 2002.  The total exposure for this analysis is $1.70\times 10^{12}$ stopped
kaons~\footnote{This is slightly less than $1.71\times 10^{12}$ stopped kaons
used for the E949 analysis~\cite{e949}.}.
The analysis is sensitive to the heavy neutrino mass region between
175 and 300~MeV/$c^2$ that corresponds to muon momentum 200 to
130~MeV/c.

\section{Experiment \label{sec:exp}}

\subsection{Detector \label{subsec:detector}}
The  E949 $K^+$ beam was produced by a high-intensity proton beam from the
Alternating Gradient Synchrotron~(AGS) at Brookhaven National 
Laboratory~(BNL).
Protons were accelerated to a momentum of 21.5~GeV/c and hit a  platinum
production target.

\begin{figure}[]
\includegraphics[width=7cm]{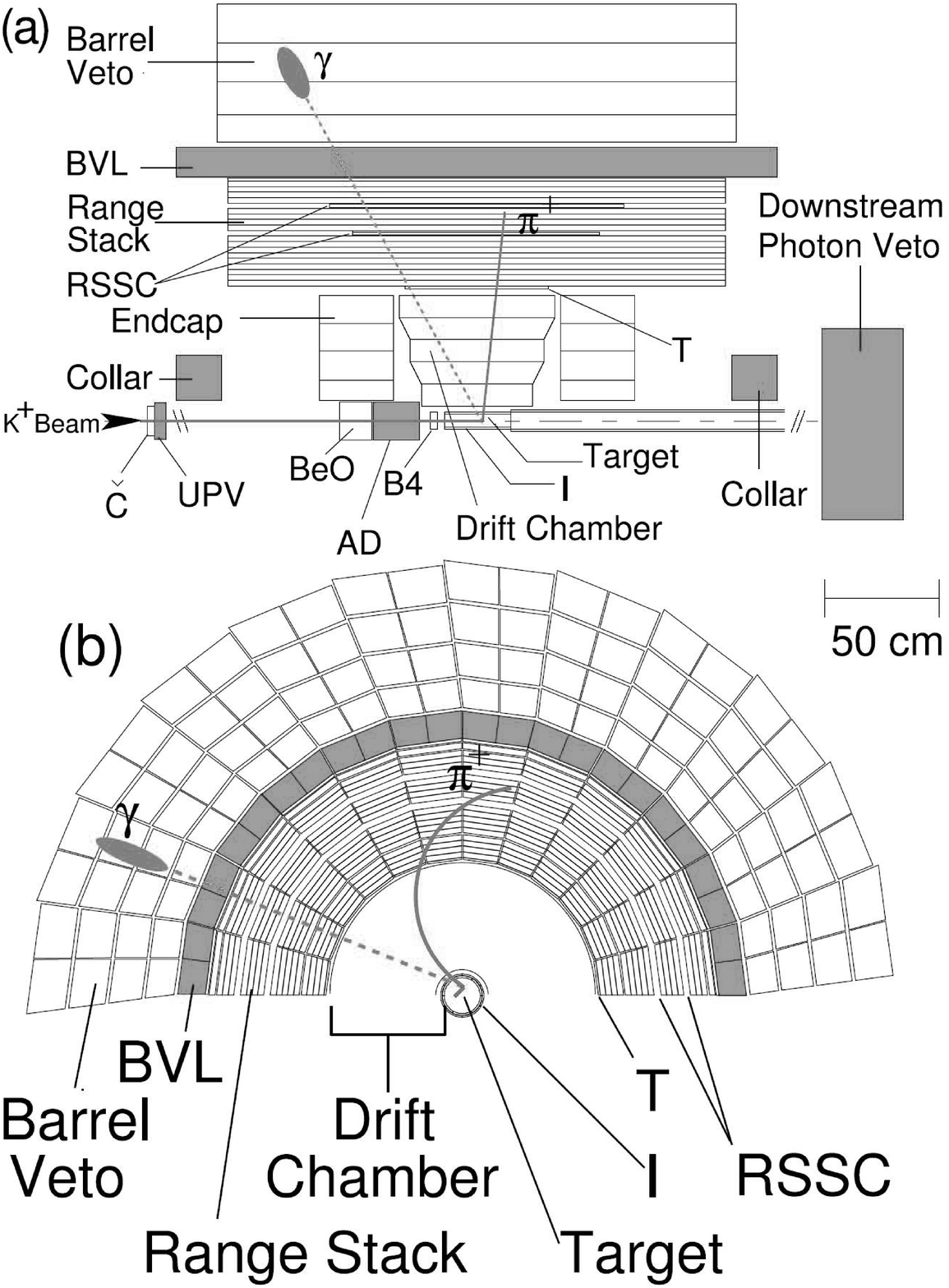}
\caption{Schematic side~(a) and end~(b) views of the upper half of the E949
detector. An incoming kaon is shown traversing the beam instrumentation,
stopping in the target, and decaying to $\pi^+\pi^0$. The outgoing charged pion
and one photon from $\pi^0\to\gamma\gamma$ decay are illustrated. Elements of
the detector are described in Section~\ref{subsec:detector}.}
\label{fig:detector}
\end{figure}

The experimental setup is illustrated in Figure~\ref{fig:detector}. Incoming
710~MeV/c kaons with  $K^+/\pi^+$  ratio of 3/1 were identified by  a
\v{C}erenkov counter.
Two beam wire chambers~(BWPCs) allowed
monitoring of the beam profile and identification of  multiple incoming
particles. Downstream of  the BWPCs, cylindrical degraders slowed the kaons so
that they came to rest in the center of the target. The inactive degrader was
made of 11.1~cm long BeO and 4.76~mm Lucite. The active degrader~(AD) 
consisted
of 40~layers of 2~mm thick scintillator disks~(139~mm diameter) alternating
with 2.2~mm thick copper disks~(136~mm diameter). The AD was split into
12~azimuthal segments. The scintillation light in each segment was sent to a
single photomultiplier tube~(PMT) through wavelength shifting fibers and read
out by analog-to-digital converters~(ADCs), time-to-digital converters~(TDCs)
and charge-coupled devices~(CCDs). Using this information the AD allowed
identification of the beam particles and detection of activity coincident
with kaon decays. After passing through the degrader, a beam hodoscope~(B4)
detected the incoming particle and identified it as a kaon by measuring the
energy deposit.

The target consisted of 413 plastic scintillating fibers 3.1~m long with a 5-mm
square cross
section to form a 12~cm diameter cylinder. A number of smaller fibers~(``edge''
fibers) filled in the gaps near the outer edge of the target. Each 5-mm fiber
was connected to a PMT, whereas the edge fibers were grouped into 12 and each
group of the edge fibers was connected to a single PMT. The PMTs were read out
by ADCs, TDCs and CCD digitizers. The fiducial region of the target was defined
by two layers of six plastic scintillation counters that surrounded the target.
The inner counters~(IC) tagged decay products for a trigger before they entered
the drift chamber. The outer counters~(VC) overlapped the downstream edge of
the IC by 6~mm and served to detect particles that decayed downstream of the
fiducial region.

The drift chamber, ``Ultra Thin Chamber''~(UTC), was located
outside of the IC. The whole E949 spectrometer was in a 1~Tesla magnetic field.
Positively charged particles were bent clockwise in the view from downstream.
The primary functions of the ~UTC were the  momentum measurement of charged
particles and providing a match between the tracks in the target and the range 
stack explained in the next paragraph.
The UTC had a length of 51~cm and inner and outer radii of 7.85~cm and 
43.31~cm,
respectively.

The range stack~(RS) was outside of the UTC at an inner radius of
45.08~cm and an outer radius of 84.67~cm. It consisted of 19 layers of plastic
scintillators azimuthally segmented into
24~sectors. The scintillators of layers 2-18 had
a thickness of 1.905~cm and a length of 182~cm. The scintillators of layer 19
had a thickness of 1~cm and were mainly used to veto charged particles with
long range by requiring that they did not reach this layer. The innermost
counters, called T-counters, served to define the fiducial volume for kaon
decay
products. The scintillation light was transmitted by light guides to PMTs. Each
PMT was read out by an ADC, a TDC and a transient digitizer (TD). The primary
functions of the RS were energy and range measurements of charged particles 
and
their identification.

The detection of any activity coincident with the charged track is very
important for suppressing the backgrounds for $K^+\to\mu^+\nu_H$ decay.
Photons from $K_{\pi 2}$ and other radiative decays were detected by hermetic
photon detectors with $4\pi$ solid angle coverage. Vetoing photons was
accomplished using the Barrel Veto~(BV), the Barrel Veto Liner~(BVL), the
upstream and downstream End Caps~(ECs), the upstream and downstream Collar
detectors~(CO), the downstream Microcollar detector~(MC), as well as the
target, RS and AD. The BV and BVL with a thickness of 14.3 and 2.29 radiation
lengths at normal incidence, respectively, provided photon detection
over
2/3 of $4\pi$ solid angle. The photon detection over the remaining 1/3 of
$4\pi$ solid angle was provided by the other calorimeters in the region from
$10^\circ$ to $45^\circ$ of the beam axis with a total thickness from 7 to
15~r.l.

The coordinate system of the detector is defined such that the origin is at the
center of the target; the $z-$axis is along the beam direction; and the
$x-$axis and $y-$axis are set in the horizontal and vertical directions,
respectively.

A more detailed description of the E949 experiment can be found in~\cite{e949}.

\subsection{Monte Carlo} \label{subsec:MC}
The detector and the physics processes in it were modeled by the 
electromagnetic-shower simulation package EGS4 program~\cite{mc}.
The simulation of kaon
decays in the E949 detector starts from a beam file with the $x$, $y$ and $z$
positions of kaon decays in the target obtained from an analysis of the
$K^+\to\mu^+\nu_\mu$ decays.
The simulation result includes all of the detector elements,
but without the beam instrumentation upstream of the target, and
the TD and CCD pulse-shape information.

\subsection{Trigger \label{subsec:trigger}}

The experimental signature of the $K^+\to\mu^+\nu_H$ decay is similar to
the $K^+\to\pi^+\nu\bar{\nu}$ decay: one single charged track with no other
detector activity. This motivates the use of the main E949 trigger to search
for heavy neutrinos. It consists of several requirements:
\begin{itemize}
    \item $K^+$ stop requirements. A kaon
must enter the target; this was checked by coincidence of the kaon \v{C}erenkov
detector, the B4 hodoscope and the target with at least 20 MeV energy deposit.
To ensure that the kaon decays at rest,  the secondary charged particle must
hit the IC at least 1.5~ns later than the kaon hit in the \v{C}erenkov
detector.
	\item Fiducial region and range requirements on charged 
tracks~($A_{Fid\&Range}$).
A charged track from a kaon decay must enter the fiducial volume of the
detector; this was checked by coincidence hits in the IC and two first layers~
(T-Counter and layer 2) of the RS in the same sector. Low energy charged tracks
 from $K^+$ decays
were suppressed by the RS layer requirement that they must reach at least the 
sixth layer of the RS.
    \item Long tracks~(in general, muons from $K_{\mu 2}$ decay) were 
    suppressed by the
layer 19 veto
requirement~($A_{\overline{19_{ct}}}$).
	\item Online pion identification~($A_{\pi\to\mu}$). It
required a signature of $\pi^+\to\mu^+$ decay in the online-selected  stopping
counter. The $\mu^+$ from the $\pi^+\to\mu^+\nu_\mu$ decay at rest has the
kinetic energy of 4~MeV~(a few~mm equivalent range in plastic scintillator) and
rarely exits the stopping counter. So, pion pulses in the stopping counter
recorded by the TDs have a double-pulse structure.
Despite the online pion identification requirement, some muons remained in the
final sample due to inefficiency.
	\item Refined range requirements on a charged track~($A_{RefinedRange}$).
It took into account the number of target fiber
hits and the track's
downstream position~($z$-coordinate) in RS layers 3, 11, 12, 13 as well as the
deepest layer of penetration. This condition suppressed long muon tracks
which passed the layer 19 veto requirement.
	\item Online photon veto. Events were rejected if any activity in the BV,
BVL or EC with energy above a
threshold was detected. This condition removed events with photons. A similar
requirement in the RS is also applied. The 24 sectors of the RS are
conventionally grouped into six; a group of 4 sectors is called a "hextant".
Only one hextant is allowed to have hits or two hextants if they are adjacent.
This rejects events with multiple tracks and events with photon activity in the
RS.
\end{itemize}

\begin{figure}[]
\centering
\includegraphics[width=\columnwidth]{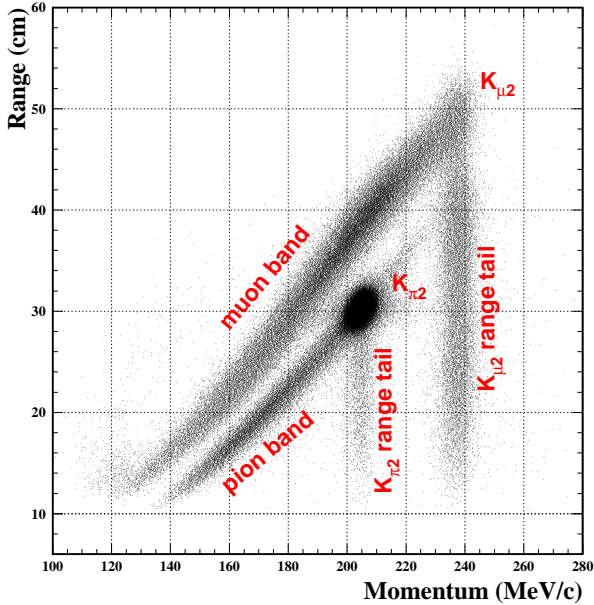}
\caption{Range in  plastic scintillator vs. the momentum of the charged
particles for events that pass the main E949 trigger. The concentration of
events due to the two-body decays are labeled $K_{\mu 2}$  and $K_{\pi 2}$.
The decays $K^+\to\mu^+\nu_\mu\gamma$ and $K^+\to\pi^0\mu^+\nu_\mu$ 
contribute
to the muon band. The pion band contains the $K^+\to\pi^+\pi^0\gamma$ decays,
$K_{\pi 2}$ decays in which the $\pi^+$ scattered in the target or range stack
and beam $\pi^+$ that scattered in the target. Both the $K_{\pi 2}$ range tail
and the $K_{\mu 2}$ range tail contain events from $K_{\pi 2}$ and $K_{\mu 2}$
decays, respectively, due to elastic~(or inelastic) scattering in the range
stack.}
\label{fig:sample}
\end{figure}

A sample of data selected by the main E949 trigger is shown in
Figure~\ref{fig:sample}, where $K_{\mu 2}$, $K_{\pi 2}$, pion band, muon band,
$K_{\mu 2}$ range tail, and $K_{\pi 2}$ range tail are defined.

In addition to the main trigger, various monitor triggers also collected events
for use in data quality assessments, calibrations of the detector subsystems
and acceptance calculations~\cite{e949_tr}. In our analysis we used 
K$\mu$2, K$\pi$2(1),
$\pi_{scat}$ and $Kbeam$ monitor triggers.

The K$\mu$2 trigger was designed to collect
muons from the $K^+\to\mu^+\nu_\mu$ decay. Since the final state does not
contain photons or additional tracks, it is a convenient sample to study
beam and target related acceptance factors described in next section for our
signal $K^+\to\mu^+\nu_H$ decay. The K$\mu$2 trigger requirements are the
following: a kaon must enter the target; the outgoing charged track must  reach
the sixth or seventh layer of the RS and then hit the 17th, 18th or 19th layer of
the RS.

The K$\pi$2(1) trigger was designed to collect pions from the
$K^+\to\pi^+\pi^0$ decay. The requirements are the following: a kaon must enter
the target; the outgoing charged track must reach at least the sixth layer of the
RS and must not hit 19th layer of the RS. According to these requirements muons
from the $K^+\to\mu^+\nu_\mu\gamma$ or $K^+\to\mu^+\pi^0\nu_\mu$ decays 
also
can pass the K$\pi$2(1) trigger. We extracted these muons to study
acceptance factors such as online pion identification and range-momentum
consistency~(described below).

Among the incoming beam particles there were many pions, including some
scattered into the fiducial volume of the RS. These scattering events were
selected by special $\pi_{scat}$ trigger to study track quality. It requires
that a pion enters the target and has an in-time track in the RS.

In addition, the trigger for beam kaons, $Kbeam$, was also defined to study
trigger efficiency and detector geometrical alignment. It requires that a kaon
enters the target.

In the analysis, common acceptance factors about kaon beam, track quality, etc.
measured by $\pi_{scat}$ and $Kbeam$ triggers
are taken from a previous analysis~\cite{e949} (more detail in later sections).

\section{Analysis \label{sec:analysis}}

\subsection{Strategy \label{subsec:strategy}}

The method used for identification of heavy neutrinos is to search for
additional peaks below the main $K_{\mu 2}$ peak. To do such a measurement we
need to know the total acceptance for the $K^+\to\mu^+\nu_H$ decay channel
and signal shape.

The total acceptance was measured using muon samples taken by the K$\mu$2,
K$\pi$2(1) triggers and Monte Carlo simulation. The total acceptance was
verified by determining the branching ratios of the
$K^+\to\mu^+\nu_\mu$ decay~(high momentum region) and the
$K^+\to\mu^+\nu_\mu\gamma$ decay~(low momentum region).

The signal shape was studied using Monte Carlo simulation of the
$K^+\to\mu^+\nu_H$ decay.

The full E949 data sample was split into 1/20
and 19/20 samples. The 1/20 sample was selected by choosing every twentieth
event.  The 1/20 data sample was used to study background shapes, systematic
uncertainties and to verify the total acceptance measurement. The 19/20 sample 
was kept separate and not accessed until all cuts were determined.

\subsection{Offline selection criteria \label{subsec:cuts}}

In this analysis we used eight groups of offline selection criteria.
\begin{enumerate}
    \item Track reconstruction in the UTC~(UTCQUAL). It rejected events with
poor UTC fits in either $x-y$ or $z$. Events with overlapping tracks were also
rejected.
    \item Kinematic cuts were used to select events in the fiducial volume of
the detector. It was required that the  cosine of the polar angle of a charged
track~($\cos\theta$) was within~$\pm 0.5$~($|\cos\theta|\leq 0.5$) and matched
the $z$ stopping location determined for  each range stack layer. Also, the
charged track must pass the UTC fiducial volume cut. The $z$ position at the
UTC
outer layer should be within the active region~($|z|\leq 25$~cm).
		\item RS track reconstruction and matching with the UTC track~(PRRF).
This cut rejected events with the charged track scattering in the RS.
	 \item Beam cuts. Several cuts were used to identify the incoming
particle as a  kaon and remove extra beam particles at the track
time~\cite{e949}.
	 \item Delayed coincidence~(DELC). This cut removed kaon
decays-in-flight by requiring that a kaon should decay at least 2~ns after it
entered the target.
	 \item Target cuts. Numerous requirements were placed on the activity
in the target to suppress random background and ensure reliable 
determination of the
kinematic properties of the charged muon. These requirements were based on the
results of the reconstructed energy and time of the pion and kaon fibers, the
pattern of kaon and pion fibers relative to information from the rest of the
detector and the results of the target-track fitter~\cite{e949}.
   	 \item Range-Momentum cut~(RNGMOM). In the main E949
analysis~\cite{e949}, this cut was designed to check whether the range of a
charged track is consistent with that for pions~(pion band in
Figure~\ref{fig:sample}). In this analysis RNGMOM cut was inverted and
changed
to select the muon band.
	 \item Photon veto cuts. This cut removed events with photon activity
in the detector. For the heavy neutrino analysis we used loose and tight photon
veto cuts. The loose photon veto was used to measure the
$K^+\to\mu^+\nu_\mu\gamma$ branching ratio, the total acceptance
systematic error and study background shapes.  The tight photon veto cuts were
used
for the total acceptance estimation, to process the 19/20 sample, and for
 producing the
final result.

\end{enumerate}

\begin{figure}[]
\centering
\includegraphics[width=\columnwidth]{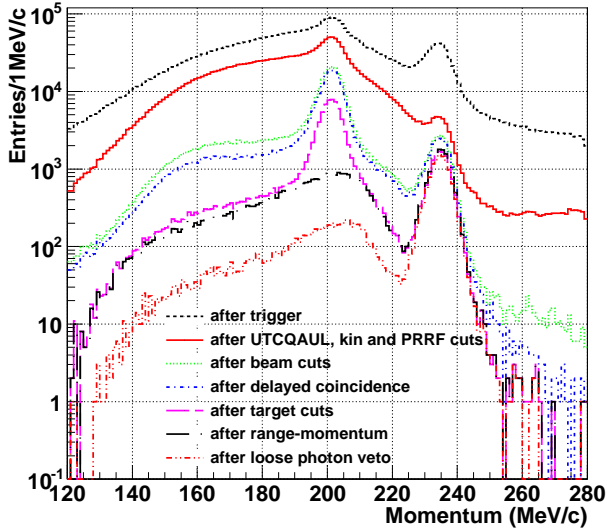}
\caption{Momentum spectra based on the 1/20 data sample
after applying each group of cuts.}
\label{fig:real}
\end{figure}

The momentum spectra based on the 1/20 data sample after applying 
each group of
cuts are shown in Figure~\ref{fig:real}.
After the beam and DELC cuts, the kaon decay-in-flight backgrounds were greatly
suppressed. After the RNGMOM cut, the  pion background was removed and the
$K_{\pi 2}$ peak disappeared. It should be noted that the RNGMOM cut was
applied only for the  muon band events with $p<220$~MeV/c. The photon veto cut
further suppressed photon backgrounds like $K^+\to\mu^+\nu_\mu\gamma$ and
$K^+\to\pi^0\mu^+\nu_\mu$. It also changed the relative magnitude of $K_{\mu
2}$ peak and the muon band.

\subsection{Total acceptance \label{subsec:TotAcc}}

\subsubsection{Measurement \label{subsubsec:meas}}
The total acceptance for the $K^+\to\mu^+\nu_H$ decay was dominated by two
main factors: online trigger requirements and offline selection criteria.
The online trigger requirements are described in Sec.~\ref{subsec:trigger} and
the offline criteria are described in Sec.~\ref{subsec:cuts}.

We used the Monte Carlo simulation to measure the acceptance of simple cuts:
the online fiducial region and range requirements on charged tracks, layer 19 veto,
refined range~ and kinematic cuts.
But the simulation package is not suitable for
some online requirements such as pion identification or delayed coincidence
because these cuts were not reliably implemented in the Monte Carlo.
Their acceptances were extracted from data.

To study the acceptance of the beam,
target, DELC, and photon veto cuts we used muons from the
$K^+\to\mu^+\nu_\mu$ decay that satisfied the K$\mu$2 trigger requirements
since this decay had the same signature as for signal $K^+\to\mu^+\nu_H$
decay: one charged track and nothing else. These cuts were assumed to be
momentum independent.

For the online pion identification and RNGMOM cut we measured the momentum
dependent acceptance curve using muons that survived the K$\pi$2(1) trigger
requirements.

The acceptances of UTCQUAL and PRRF cuts were measured using 
scattering pions
from the $\pi_{scat}$ trigger and the values were taken from the main E949
analysis~\cite{e949}.

Two factors
were also taken into account for the total acceptance estimation. The first
one, $\epsilon_{T\bullet 2}$, accounts for the acceptance loss due to the
geometrical and counter inefficiencies of the T-counters; it was measured
using $Kbeam$ trigger and its value was taken from the main E949
analysis~\cite{e949}.
The second one,
$f_s$, is the $K^+$ stop efficiency in the target; it was measured using the
K$\pi$2(1) trigger. Both $\epsilon_{T\bullet 2}$ and $f_s$ were assumed to be
momentum independent.

\begin{figure}[]
\centering
\includegraphics[width=\columnwidth]{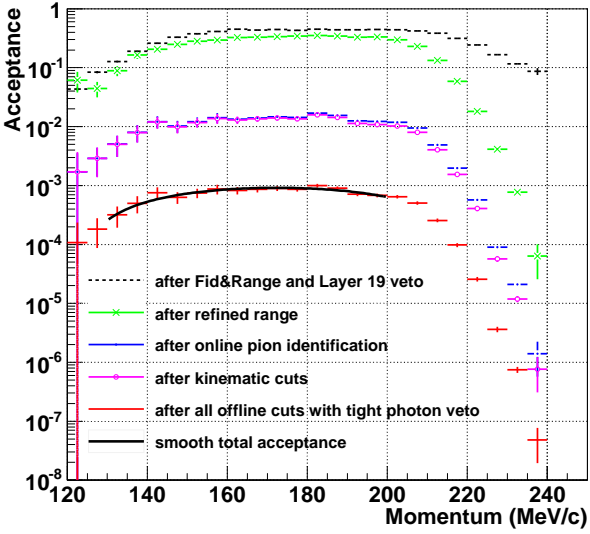}
\caption{Acceptance dependence on momentum. Black solid line shows the
smooth total acceptance which is used for the mixing matrix element upper
limit calculation.}
\label{fig:acc}
\end{figure}

The total acceptance after applied cuts is shown in Figure~\ref{fig:acc}.
The acceptance drop off below 140~MeV/c is due to the requirement that
charged track must reach at least the sixth layer of the range stack. 
The acceptance
drop off above 200~MeV/c is due to two requirements. First, the charged track
must not reach layer 19 of the range stack and second, the refined range
removes long tracks which are dominant at high momentum for events passing the
layer 19 requirement. The main acceptance loss~(factor $\sim 20$) comes from
the online pion identification requirement~(blue curve in
Figure~\ref{fig:acc}). It should be noted that the  acceptance curves in
Figure~\ref{fig:acc} must be corrected for momentum $p>220$~MeV/c since the
Monte Carlo simulation of layer 19 and the refined range requirements are  not
accurate in this region. However, this is not relevant for the heavy neutrino
study since we will investigate the momentum region between 130~MeV/c and
200~MeV/c. According to Figure~\ref{fig:acc} the acceptance is smooth and
has a maximum in this region. The total acceptance for the $K^+\to\mu^+\nu_H$
decay with heavy neutrino mass
$m_{\nu_H}=250$~MeV/$c^2$~($P_{\mu}=163.6$~MeV/c) was measured to be

\begin{equation}
    A_{m_{250}}=(8.00\pm 1.05(stat.))\times 10^{-4},
\end{equation}
where the error is statistical. The systematic uncertainty will be presented
below. So, the single event sensitivity~(S.E.S.) for the heavy neutrino with
mass $m_{\nu_H}=250$~MeV/$c^2$ can be calculated as

\begin{equation}
    S.E.S.=\frac{1}{Acc\times N_K}=7.35\times 10^{-10},
\end{equation}
where $Acc$ is the total acceptance and $N_K$ is the total number of stopped
kaons. This sensitivity is roughly constant for the whole investigated region.

\subsubsection{Verification \label{subsubsec:TotAccVer}}

The $K^+\to\mu^+\nu_\mu$ and $K^+\to\mu^+\nu_\mu\gamma$ branching 
ratios were
measured using the 1/20 data sample to verify the total acceptance for the
signal decay. Table~\ref{tab:verify} shows they share a lot of
common data samples for acceptance measurement of the signal channel.

\begin{table}[]

\begin{tabular}{|l|l|c|c|c|} \hline
\multicolumn{2}{|l|}{Acceptance factors}& $K^+\to\mu^+\nu_H$&
$K^+\to\mu^+\nu_\mu\gamma$&
$K^+\to\mu^+\nu_\mu$\\ \hline

\multicolumn{2}{|l|}{$f_s$}&\multicolumn{2}{c}{K$\pi$2(1)}&\\ \cline{1-4}
\multicolumn{2}{|l|}{$A_{PRRF}$}&
\multicolumn{2}{c|}{\multirow{2}{*}{$\pi_{scat}$}}&
 \\ \cline{1-2}\cline{5-5}
\multicolumn{2}{|l|}{$A_{UTCQUAL}$}&\multicolumn{3}{c|}{}\\ \hline

\multicolumn{2}{|l|}{$\epsilon_{T\bullet 2}$}&\multicolumn{2}{c}{$Kbeam$}&
 \\ \hline

\multicolumn{2}{|l|}{$A_{Fid\&Range}$}&\multicolumn{3}{c|}{\multirow{2}{*}
{}}
\\ \cline{1-2}
\multicolumn{2}{|l|}{$A_{Kinematic}$} & \multicolumn{2}{c}{MC}&\\ \cline{1-2}
\cline{5-5}

\multicolumn{2}{|l|}{$A_{\overline{19_{ct}}}$}&\multicolumn{2}{c|}{\multirow{2}{*}{}}
&K$\mu$2
\\ \cline{1-2}\cline{5-5}
\multicolumn{2}{|l|}{$A_{RefinedRange}$}&\multicolumn{2}{c|}{} &
\multirow{2}{*}{K$\pi$2(1)}
  \\ \cline{1-4}
\multicolumn{2}{|l|}{$A_{\pi\to\mu}$}& \multicolumn{2}{c|}
{\multirow{2}{*}{K$\pi$2(1)}}&\\ \cline{5-5}

\multicolumn{2}{|l|}{$A_{RNGMOM}$}&\multicolumn{2}{c|}{}&not applied
\\ \hline
\multicolumn{2}{|l|}{$A_{Beam\&Target}$}&\multicolumn{2}{c}
{\multirow{2}{*}{K$\mu$2}}&

 \\ \cline{1-2}
\multicolumn{2}{|l|}{$A_{DELC}$}&\multicolumn{3}{c|}{} \\
\cline{1-2}\cline{1-4}
\multirow{2}{*}{$A_{PV}$}&Loose &not applied &MC &\\ \cline{2-5}
& Tight&K$\mu$2 &not applied &not applied\\ \hline
\end{tabular}

\caption{Acceptance factors for signal decay and for the $K^+\to\mu^+\nu_\mu$
and $K^+\to\mu^+\nu_\mu\gamma$ branching ratios measurements.
Each cell indicates the data sample which was used to study corresponding
acceptance factor. MC means Monte Carlo simulation.}

\label{tab:verify}
\end{table}

The $K^+\to\mu^+\nu_\mu$ decay is similar to the signal $K^+\to\mu^+\nu_H$
decay, but it has higher muon momentum than our signal region. That's why
layer 19 veto, refined range and online pion identification acceptance
factors must be remeasured.

The $K^+\to\mu^+\nu_\mu\gamma$ decay has muons with momentum in 
signal region,
but there is an extra photon in the final state. That's why photon veto
acceptance must be studied separately and a different method was used to
measure it.

\begin{itemize}

\item $BR(K^+\to\mu^+\nu_\mu\gamma)$ measurement

The $K^+\to\mu^+\nu_\mu\gamma$ branching ratio was measured in 
the momentum
region
$140<p_\mu<200$~MeV/c.  All acceptance factors except photon veto 
were already
measured and shown in Figure~{\ref{fig:acc}}. The photon veto acceptance for
the $K^+\to\mu^+\nu_\mu\gamma$ decay should be studied separately due to the
presence of one photon in final state.

Photons from the $K^+\to\mu^+\nu_\mu\gamma$ decay have energy 
 $E>20$~MeV in
the muon momentum region $140<p_\mu<200$~MeV/c.
The photon veto acceptance in this region was determined using the
single photon inefficiency table~(see Appendix~\ref{subsec:spi} 
and~\cite{kentaro}). 
The photon acceptance was measured to be 
$A_{PV}=(1.24\pm 0.38)\times 10^{-2}$.

The total acceptance for the $K^+\to\mu^+\nu_\mu\gamma$ decay 
was determined to
be $A_{K_{\mu\nu\gamma}}=(3.60\pm 1.11)\times 10^{-5}$. Based the on 1/20
sample,
the branching ratio was measured to be

\begin{equation}
\begin{array}{ll}
    BR(K^+\to\mu^+\nu_\mu\gamma,140<p_\mu<200~\textnormal{MeV/c})= \\
    =(1.3\pm 0.4)\times 10^{-3},
\label{eq:km2g}
\end{array}
\end{equation}
where the uncertainty includes both systematic and statistical effects. The
uncertainty is completely dominated by the uncertainty in the photon
acceptance.

The Particle Data
Group~(PDG) average value of the $K^+\to\mu^+\nu_\mu\gamma$ 
branching ratio is
$(6.2\pm 0.8)\times 10^{-3}$ for $p_\mu<231.5$~MeV/c~\cite{pdg}. 
To compare our
measurement with this value we need to determine the ratio
$N_{140<p<200}/N_{p<231.5}$. This ratio was calculated from Monte Carlo
simulation of the $K^+\to\mu^+\nu_\mu\gamma$ decay with a threshold
$E_\gamma>5$~MeV to be $N_{140<p<200}/N_{p<231.5}=0.2231\pm
 0.0004(stat.)$.
Therefore, the estimated PDG value for $140<p_\mu<200$~MeV/c would be

\begin{equation}
\begin{array}{ll}
    BR^{PDG}(K^+\to\mu^+\nu_\mu\gamma,140<p_\mu<200~\textnormal{MeV/c})
    = \\
    =(1.4\pm 0.2)\times 10^{-3}
\label{eq:km2g_pdg}
\end{array}
\end{equation}

Our measurement is consistent with  the estimated PDG average within the
uncertainty.

\item  $BR(K^+\to\mu^+\nu_\mu)$ measurement

As mentioned in Sec.~\ref{subsubsec:meas}, some acceptance factors for the
momentum region
$p_\mu>220$~MeV/c require correction. This momentum region is 
crucial for the
$K^+\to\mu^+\nu_\mu$ branching ratio calculation because the mean 
muon momentum
is $p_\mu=236$~MeV/c.
Three main differently factors including layer 19, refined range and
online pion identification were remeasured sequentially using experimental
data from the K$\mu$2 and K$\pi$2(1) triggers.

The PRRF acceptance was also measured differently
since the aim of this cut is to suppress scattering in the RS and
the $K^+\to\mu^+\nu_\mu$ sample includes scattering
events~($K_{\mu 2}$ range-tail in Figure~\ref{fig:sample}). The range-momentum
cut~(RNGMOM) was not applied for the $K^+\to\mu^+\nu_\mu$ branching ratio
measurement because in our analysis we use this cut only for charged tracks
with momentum below 220~MeV/c.

The total acceptance for the $K^+\to\mu^+\nu_\mu$ decay was measured to be
$A_{K_{\mu 2}}=(1.60\pm 0.45)\times 10^{-7}$, where the uncertainty is
statistical and the main contribution comes from the refined range and online
pion identification measurement due to low statistics in the K$\pi$2(1)
trigger after these cuts were applied. Statistics are low because the
refined
range and online pion identification were designed to reject muons.
The result for the $K^+\to\mu^+\nu_\mu$ branching ratio is
$BR(K^+\to\mu^+\nu_\mu)=0.54\pm 0.15$, where the uncertainty is statistical.
This value is consistent with the world average value from PDG~\cite{pdg} ---
$BR(K^+\to\mu^+\nu_\mu)=0.6355\pm 0.0011$. The decay
$K^+\to\mu^+\nu_H$ will be studied with $p_\mu<200$~MeV/c and in this region
the refined range and online pion acceptance were measured well~(see
Figure~\ref{fig:acc}).

\end{itemize}

\subsubsection{Summary \label{subsubsec:summary}}
The summary of our acceptance study is presented in Table~\ref{tab:summary}.

\begin{table}[]
\centering
\begin{tabular}{|l|r|} \hline
 &$K^+\to\mu^+\nu_H$, \\
 &$m_{\nu_H}=250$~MeV/$c^2$\\
\hline
$N_K$& $1.70\times 10^{12}$  \\ \hline
$\epsilon_{T\bullet 2}$& $0.9505\pm 0.0012$ \\ \hline
$f_s$& $0.7558\pm 0.0075$ \\ \hline
$A_{Fid\&Range}$& $0.4383\pm 0.0011$ \\ \hline
$A_{\pi\to\mu}$& $0.0412\pm 0.0053$\\ \hline
$A_{RefinedRange}$ & $0.7252\pm 0.0159$\\ \hline
$A_{UTCQUAL}$& $0.9503\pm 0.0007$\\ \hline
$A_{Kinematic}$& $0.9662\pm 0.0006$ \\ \hline
$A_{PRRF}$& $0.9520\pm 0.0007$\\ \hline
$A_{Beam\&Target}$& $0.5102\pm 0.0003$\\ \hline
$A_{DELC}$& $0.7672\pm 0.0002$\\ \hline
$A_{RNGMOM}$& $0.9739\pm 0.0012$\\ \hline
$A_{PV}$& $0.2551\pm 0.0012$\\ \hline \hline
$A_{total}$& $(8.00\pm 1.05(stat.)\pm 2.46(syst.))\times 10^{-4}$ \\ \hline
\hline
&$S.E.S.=7.35\times 10^{-10}$\\ \hline

\end{tabular}
\caption{Summary for the acceptance measurement of the 
$K^+\to\mu^+\nu_H$ decay
with heavy neutrino mass $m_{\nu_H}=250$~MeV/$c^2$.}
\label{tab:summary}
\end{table}
The errors in this table for individual cuts are statistical, and the
systematic error for the total
acceptance is from the error on the $K^+\to\mu^+\nu_\mu\gamma$ branching ratio
measurement.

\subsection{Residual background \label{subsec:bkg}}

The search for $K^+\to\mu^+\nu_H$ seeks evidence for  additional peaks below
the $K_{\mu 2}$ peak. So, all background sources that could mimic the  signal
must be understood. We simulated the main background sources,
$K^+\to\mu^+\nu_\mu\gamma$, $K^+\to\pi^0\mu^+\nu_\mu$ and
$K^+\to\pi^+\pi^0\gamma$ decays. After the trigger requirements and
offline selection criteria, the $K^+\to\pi^0\mu^+\nu_\mu$ contribution in the
total number of background events is less than 1\% of the
$K^+\to\mu^+\nu_\mu\gamma$ contribution due to the presence of two photons in
the final state. The $K^+\to\pi^+\pi^0\gamma$ decay can be ignored due to three
photons in the final state and the large range-momentum pion rejection~(which
removes the pion band in Figure~\ref{fig:sample}). Therefore, the
$K^+\to\mu^+\nu_\mu\gamma$ is the dominant background source in 
the search for
$K^+\to\mu^+\nu_H$ decays.

Given the agreement between the PDG values and our 
$K^+\to\mu^+\nu_\mu$ and
$K^+\to\mu^+\nu_\mu\gamma$ branching ratio measurements, the
experimental muon momentum spectra~(see dashed-double dotted line in
Figure~\ref{fig:real}) and the simulated $K_{\mu 2}+K_{\mu\nu\gamma}$ muon
momentum spectra can be compared.

\begin{figure}[]
\centering
\includegraphics[width=\columnwidth]{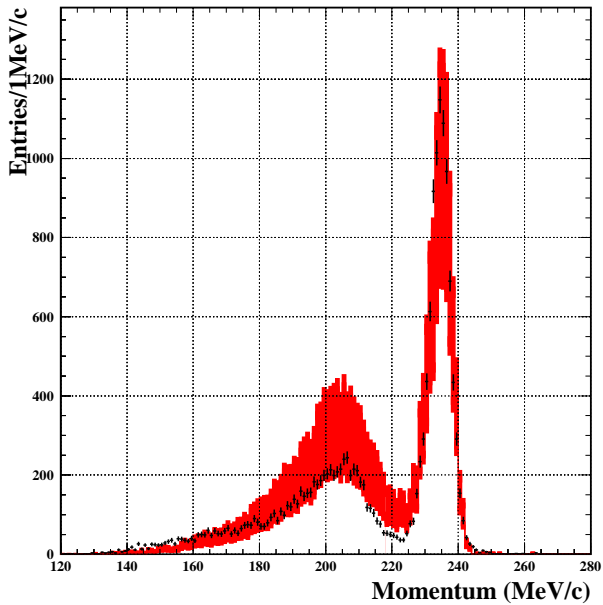}
\caption{Momentum spectra for simulated $K_{\mu 2}$+$K_{\mu\nu\gamma}$
events and experimental events based on the 1/20 data sample. The shaded band
shows the $\pm 1\sigma$ spread with known acceptance uncertainties.
The black dots are the 1/20 data. Colors are available online.}
\label{fig:finalexpmc}
\end{figure}

The momentum spectra for the simulated 
$K_{\mu 2}$+$K_{\mu\nu\gamma}$ events
and experimental events based on the 1/20 data sample are shown in
Figure~\ref{fig:finalexpmc}, where $K_{\mu 2}$ and $K_{\mu\nu\gamma}$
were normalized according to their branching ratios. 
The red band shows the $\pm 1\sigma$ spread with
the known acceptance uncertainties. The $K_{\mu\nu\gamma}$ central histogram
uses $A_{PV}= 1.27\times 10^{-2}$.

The momentum spectrum for data and MC do not agree.
Between 200 MeV/c and 220 MeV/c, the radiative gamma energy is low. The
difference is caused by the difficulty in simulating detector activity  or
electronic noise of the low photon veto cut threshold. Beyond 220 MeV/c, it is
caused by the uncertainty of layer 19 and refined range cuts.
Below 200 MeV/c, the trends of simulated and experimental spectra 
are consistent.

Since the simulated shape does not show obvious bumps or valleys, we assume
that the experimental background shape is also smooth.

\subsection{Resolution \label{subsec:res}}

\begin{figure}[]
\centering
\includegraphics[width=\columnwidth]{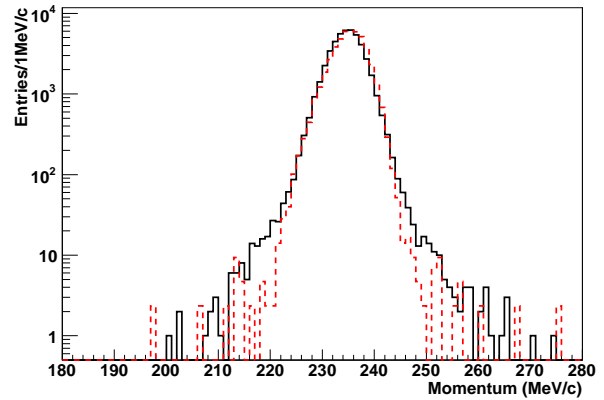}
\caption{Simulated $K^+\to\mu^+\nu_\mu$~(dashed red) and
experimental decays derived from K$\mu$2 trigger~(solid black). Colors online.}
\label{fig:reskm2}
\end{figure}

\begin{figure}[]
\centering
\includegraphics[width=\columnwidth]{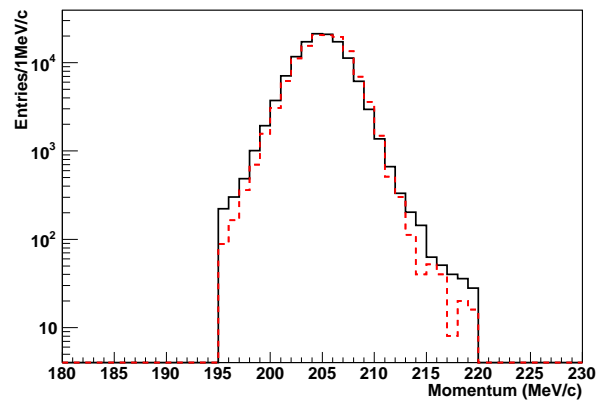}
\caption{Simulated $K^+\to\pi^+\pi^0$~(dashed red) and experimental decays
derived from K$\pi$2(1) trigger~(solid black). Colors online.}
\label{fig:reskp2}
\end{figure}

The observed momentum distribution of two-body $K^+\to\mu^+\nu_H$ 
decays would
be smeared by scattering and resolution effects. To obtain the expected shape,
we compare Monte Carlo simulations with well-known decays, $K_{\mu 2}$ and
$K_{\pi 2}$, which were derived from monitor triggers. The results are shown in
Figure~\ref{fig:reskm2} and Figure~\ref{fig:reskp2}.


Both $K_{\mu 2}$ and $K_{\pi 2}$ simulated spectra are in a good agreement with
the experimental spectra. The widths agree to within  2~\% and the mean values
agree to 
within 0.3~\%,
the tails are simulated to the $10^{-3}$ level. We concluded that 
we may rely on Monte Carlo
simulation to reproduce the detector momentum resolution.

Since the  signal region is $130<p_\mu<200$~MeV/c,   we simulated
$K^+\to\mu^+\nu_H$ decays with different heavy neutrino masses for which the
muon momentum is within the signal region.

\begin{figure}[]
\centering
\includegraphics[width=9cm]{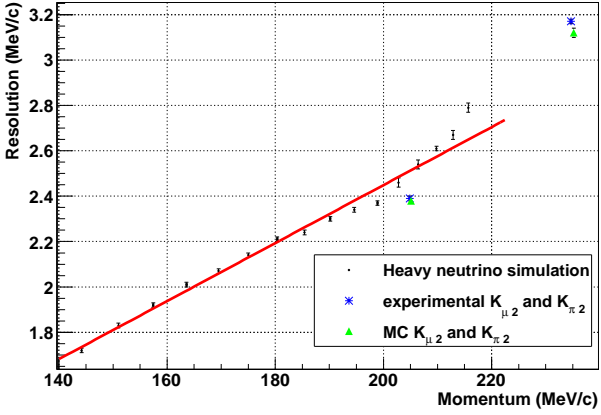}
\caption{Momentum resolution. The $K_{\mu 2}$ and $K_{\pi 2}$ points were
measured with the K$\mu$2 and K$\pi$2(1) triggers, while heavy neutrino points
were measured within the main E949 trigger. The $K_{\mu 2}$, $K_{\pi 2}$ and
heavy neutrino points cannot be comparable between each other at the same
momentum because different selection criteria were applied for each sample.}
\label{fig:res_final}
\end{figure}

The signal after all cuts is well reproduced by  a Gaussian function and we
used the standard deviation of  the fit to measure the detector resolution.
The $K^+\to\mu^+\nu_H$ decay with 16 different heavy neutrino masses within the
main E949 trigger was simulated. The momentum resolution distribution is shown
in Figure~\ref{fig:res_final}.
The linear fit gave us the expression for the resolution dependence on
momentum as

\begin{equation}
\begin{array}{ll}
    \sigma(p)(MeV/c)=(-0.1071+0.01278\times p(MeV/c))\pm \\
\pm 0.14\pm 0.05
\label{eq:res}
\end{array}
\end{equation}
where the first error is the maximum difference between the MC heavy neutrino
points and the fitted line.
The source of the second error is the maximum difference between the Monte
Carlo simulation and the data points for the decays $K_{\mu 2}$ and $K_{\pi
2}$.

\subsection{Peak search \label{subsec:method}}
To search for heavy neutrino peaks
asymptotic formulae for likelihood-based tests were used, 
which were derived using the results
of Wilks and Wald~\cite{Cowan:2010js}.
The method is a frequentist approach
which is free of computationally expensive Monte Carlo calculations and is able
to consider the shape of the signal. It thus avoids the ambiguity of selecting
a signal region~(like three or five times the resolution). 
In addition to the mean value of the upper limit, an error band of
the upper limit can be also calculated.

The following
likelihood function was used in the analysis
\begin{equation}
\begin{array}{ll}
L(\mu,\theta)=&\{\prod_{i=1}^{N_{bin}}\frac{({\mu}\cdot\epsilon
s_i+{\beta}b_i)^{n_i}}{n_i!}e^{-({\mu}\cdot\epsilon s_i+{\beta}b_i)}\}\\
&\\
&\times Gauss(\epsilon; \epsilon_{peak},\sigma_{\epsilon_{peak}}),
\label{eq:like}
\end{array}
\end{equation}
where $N_{bin}$ is the number of bins of the fitting region; $s_i$ and $b_i$ are
the probabilities of signal and background events in the $i$th bin 
after normalization, respectively;
$n_i$ is the number of observed events in the $i$th bin;
the nuisance parameter $\beta$ gives the total background strength;
the nuisance parameter $\epsilon$ is added to correct the total acceptance
via a Gaussian distribution which has a mean at 
$\epsilon_{peak}$~(Figure~\ref{fig:acc}) and a sigma at
$\sigma_{\epsilon_{peak}}$ (table~\ref{tab:summary});
and $\mu$ gives the total signal strength.
The signal shape $s_i$ is a Gaussian
distribution with standard deviation $\sigma(p)$, which is the
momentum resolution in MeV/c as a function of the momentum 
$p$~(Eq.~\ref{eq:res}).
The background shape $b_i$ and its mean strength was determined by data
and will be discussed in detail later.

A profile likelihood ratio was constructed for a signal strength parameter 
$\mu$ being tested
\begin{equation}
\lambda(\mu)=\frac{L(\mu,\hat{\hat{\theta}})}{L(\hat{\mu},\hat{\theta})},
\end{equation}
where the two nuisance parameters are represented as $\theta=(\beta, \epsilon)$;
$\hat{\hat{\theta}}$ in the numerator denotes the value of
$\theta$ that maximizes likelihood \emph{L};
the denominator is the maximized likelihood function with all parameters free and
$\hat{\mu}$ and $\hat{\theta}$ are the best fit results.
Given a set of parameters, the $\hat{\mu}$ follows a Gaussian distribution 
caused by the statistical fluctuation of the data sample and the mean value of 
$\hat{\mu}$ is the true signal strength represented by $\mu'$. 
This gives rise to the random distribution of the profile likelihood ratio.

More conveniently, the statistic
\begin{equation}
t_{\mu}=-2\ln\lambda(\mu)
\end{equation}
has a known distribution which is a chi-square distribution for one degree of 
freedom when $\mu$ equals to the true signal strength $\mu'$, for example 
$\mu$=$\mu'$,
and the most probable value of $t_{\mu}$ is zero. When $\mu$ is different with 
the true signal strength $\mu'$,
this test statistics follows a noncentral chi-square distribution for one degree of 
freedom and its median value (peak) deviates from zero, 
therefore the upper limit of $\mu'$ can be determined when the coverage at 
the value of $t_{\mu}$ meets the required confidence level based on the 
$t_{\mu}$ distribution under the condition of $\mu$=$\mu'$=0.
The error band of the upper limit can be also derived from the spread of the
 $t_{\mu}$ distribution.

The concept of $t_{\mu}$ was expanded to take into account 
the physical boundary
\begin{equation}
\label{eq:tmu}
t_{\mu}=\left\{
          \begin{array}{ll}
            -2\ln{\frac{L(\mu,\hat{\hat{\theta}})}{L(0,\hat{\hat{\theta}})}}, &
\hbox{$\hat{\mu}<0,$} \\
            & \\
            -2\ln{\frac{L(\mu,\hat{\hat{\theta}})}{L(\hat{\mu},\hat{\theta})}},
& \hbox{$\hat{\mu}\geq0.$}
          \end{array}
        \right.
\end{equation}
where the best guess of $\hat{\mu}$ was set to zero when its best fit result was
 negative,
and this test statistic gives results consistent with the 
Feldman-Cousins method~\cite{fc}.
Due to the complexity of this segmented function, the integral was calculated
numerically 
to scan the coverage of $t_{\mu}$.

According to~\cite{Cowan:2010js}, the Asimov data set ($\mu$=$\mu'$=0) 
was used to evaluate the
expected upper limit and its error band. For the heavy neutrino search in
this paper, the background shape in the Asimov data set was
determined directly by fitting the momentum spectrum of data after all
criteria.
To avoid artificial peaks or valleys in the signal region, the range
$\pm 9\sigma$ (the $\sigma$ is the momentum resolution in Eq.~\ref{eq:res})
around the point of interest was chosen to fit for background with a second
order polynomial function.
The observed limit on data was extracted with Eq.~\ref{eq:tmu}.

The 1/20 sample with loose photon veto in
Figure~\ref{fig:real} was chosen to test fit quality. The tight photon veto is
not suitable for this due to very low statistics in the 1/20 sample. The
background fitted result's $\chi^2/ndf$ varies between 0.7 and 1.4, where the
$ndf$ is the number of degrees of freedom and corresponds to the number of
points used in the fit minus the number of fit parameters.

\section{Results \label{sec:results}}
With the real data, the Asimov data and the test statistic $t_\mu$,
the mixing matrix element $|U_{\mu H}|^2$ upper limit can be obtained for a
fixed
momentum value. The mixing matrix element upper limit was calculated using the
equation below which can be derived from Eq.~\ref{eq:u}
\begin{equation}
    |U_{\mu H}|^2=\frac{N_{candidates}}{Acc\times N_K \times\rho\times
BR(K_{\mu 2})},
\label{eq:mixing_value}
\end{equation}
where $Acc$ is the total acceptance, $N_K$ is the number of stopped kaons,
$\rho$ is a kinematical factor~\cite{shrock} and $BR(K_{\mu 2})=0.6355$ is the
$K^+\to\mu^+\nu_\mu$ branching ratio~\cite{pdg}. 
According to constructed likelihood function~(Eq.~\ref{eq:like}), 
the signal strength parameter $\mu$
is not the number of candidate events itself, but number of candidate events
after correcting for acceptance. Therefore,
the value $N_{candidates}/Acc$
is the strength parameter $\mu$
and the upper limit of $\mu$ leads to the upper limit of $|U_{\mu
H}|^2$.

\begin{figure*}[]
\centering
\includegraphics[width=15cm]{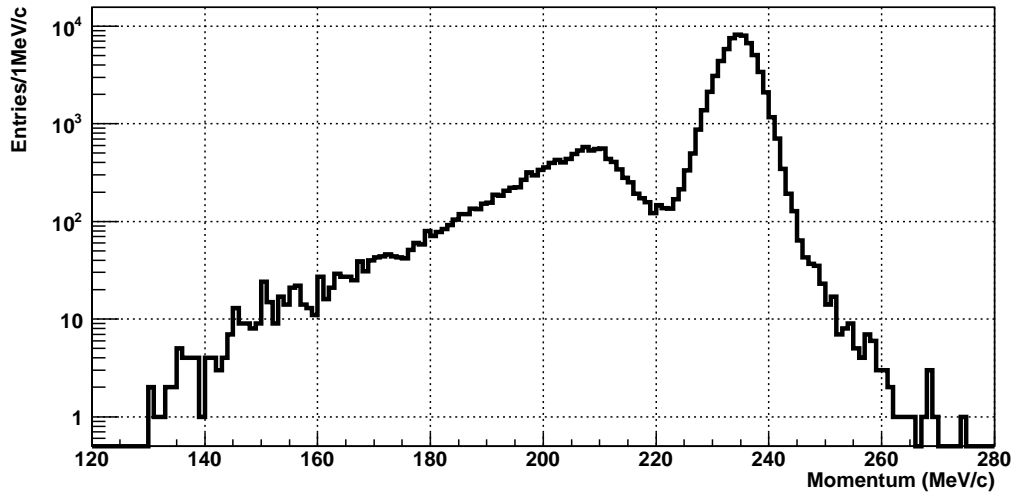}
\caption{Muon momentum spectrum for the full E949 data sample after all cuts
applied.}
\label{fig:full}
\end{figure*}

After the 1/20 data sample analysis and the peak search method was tested,
we proceeded to analyze the full E949 data sample and applied the tight PV.
The muon momentum spectrum after all cuts is shown in Figure~\ref{fig:full}.
However, after processing the 19/20 sample for peak finding, 
we found that the $\pm 9\sigma$ region was not suitable for
the  high momentum region and a $\pm 6\sigma$ region was used. 
The background
fitted results $\chi^2/ndf$ varies between 0.5 and 2.5 for the $\pm 9\sigma$
region and between 0.4 and 1.6 for the $\pm 6\sigma$ region.

\begin{figure*}[]
\centering
\includegraphics[width=15cm]{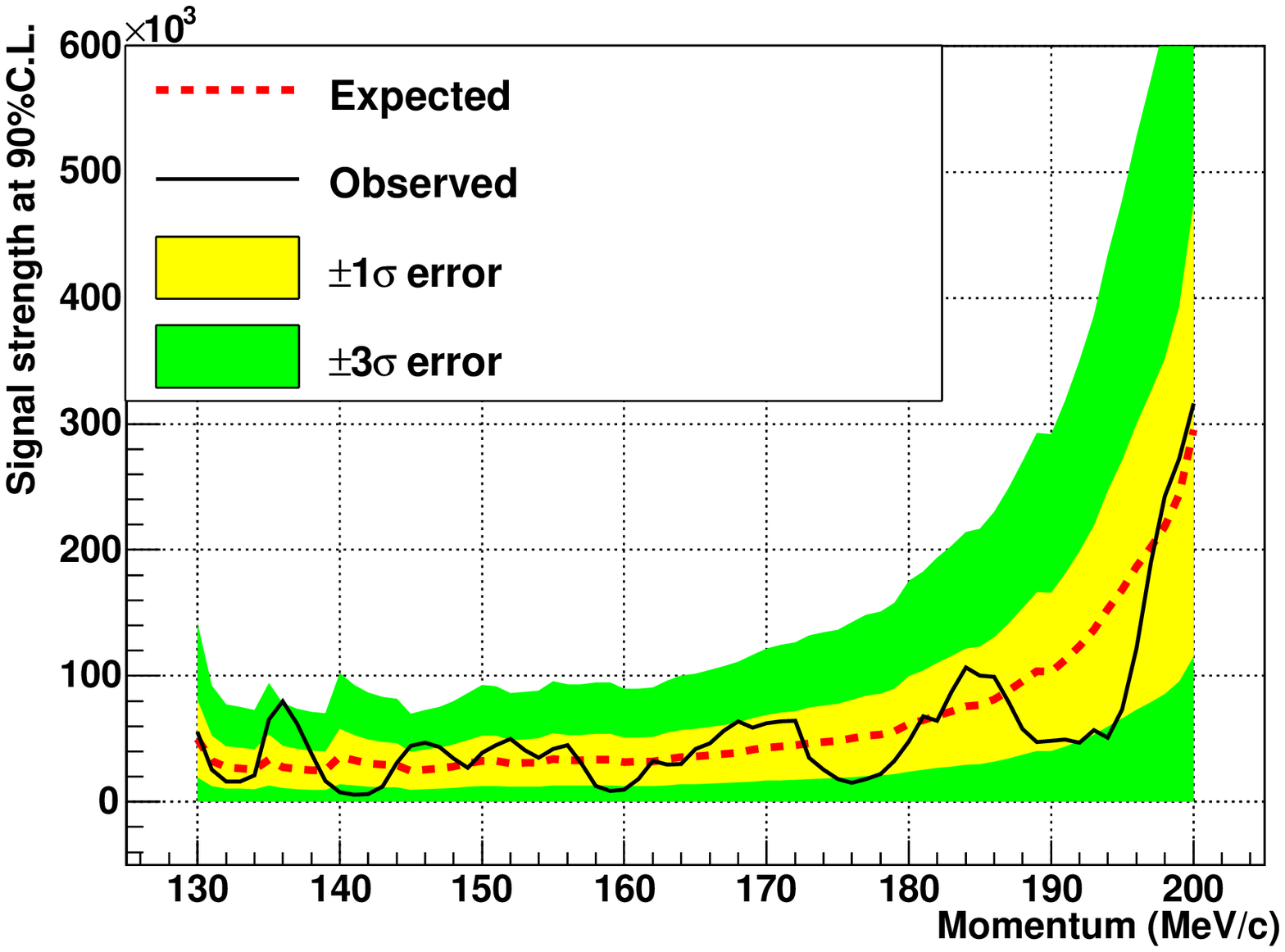}
\caption{Signal strength (defined in text) vs. muon momentum. 
The (red) dotted line is the 90\%C.L. expected upper limit with a 
$\pm1\sigma$  error
band and $\pm3\sigma$ error band. The black line is the observed upper limit
result.  Colors are
available online.}
\label{fig:result}
\end{figure*}

The signal strength parameter $\mu$ vs. muon momentum is shown
in Figure~\ref{fig:result}.
The dotted~(red) line is the expected upper limit using the Asimov data and the
solid~(black) line is the observed result using real data. In addition,
the~(color) filled area is the 1$\sigma$ and $3\sigma$ error bands
evaluated with the Asimov data in the momentum region from 130~MeV/c to
200~MeV/c.
The uncertainty of the upper limit calculation is dominated by the statistics
and the error of the total acceptance.

\begin{figure*}[]
\centering
\includegraphics[width=15cm]{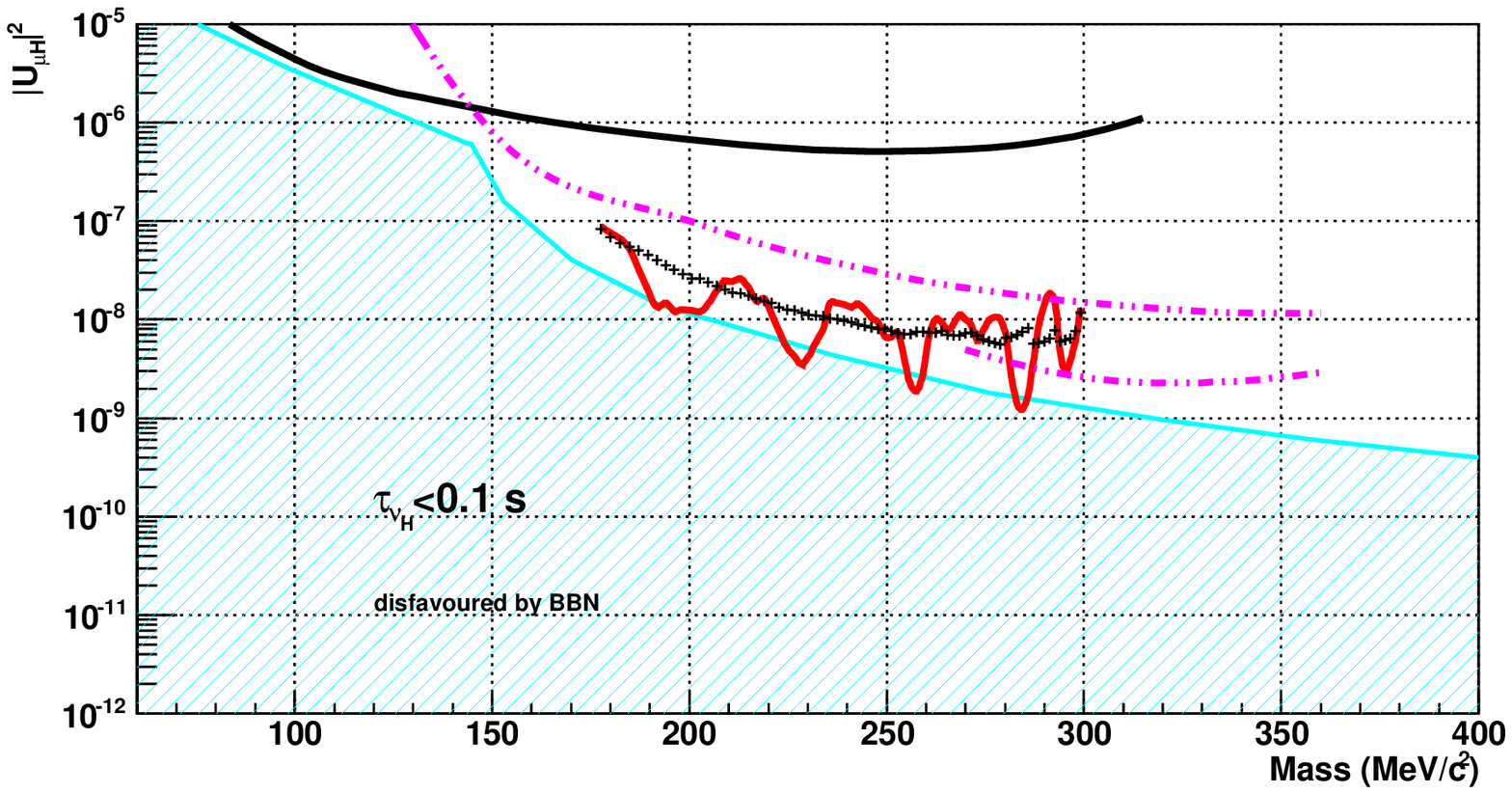}
\caption{90\% C.L. upper limits on the mixing matrix element 
$|U_{\mu H}|^2$ set by this
experiment~(solid red curve, black crosses show expected upper limit) and
others.
The solid smooth black line shows the result of a previous peak search
in kaon decays~\cite{ex1},
The dot-dash magenta lines show the results of the heavy
neutrino decay experiment CERN PS191~\cite{ps191} in two modes: the upper
dot-dash line is
derived from $K^+\to\mu^+\nu_H\to\mu^+(\mu^-e^+\nu_e)+c.c.$, 
the lower dot-dash
line is derived from $K^+\to\mu^+\nu_H\to\mu^+(\mu^-\pi^+)+c.c.$ The blue
shaded region
shows one of the possible BBN lower bounds~\cite{prediction, prediction2}. 
Colors are
available online.}
\label{fig:mixing}
\end{figure*}

There is no evidence for  a heavy neutrino signal. According to our
constructed likelihood~(Eq.~\ref{eq:like}) $\mu$ in Figure~\ref{fig:result}
(y-axis) is
$N_{candidates}/{Acc}$ at 90\%~C.L. and we can use it
directly to calculate the mixing matrix element upper
limit~(Eq.~\ref{eq:mixing_value}).
This result for the mixing matrix element
upper limits at 90\%~C.L. is shown in Figure~\ref{fig:mixing} varying from
$10^{-9}$ to $10^{-7}$.

\section{Summary}
We reported the result of the search for heavy neutrinos in the
$K^+\to\mu^+\nu_H$ decay channel using the E949 data sample in an 
exposure of
$1.70\times 10^{12}$ stopped kaons. Heavy neutrinos with
masses $\mathcal{O}(1)$~GeV/$c^2$ are allowed by the $\nu$MSM model.  
The main
E949 trigger was designed to select pions, but muons were present in data set
due to inefficiencies in the pion selection criteria applied. These muons were
used for the search for heavy neutrinos. Since no evidence for extra peaks
below the main $K^+\to\mu^+\nu_\mu$ peak was found we set 
new upper bounds on
the mixing matrix element $|U_{\mu H}|^2$ in the mass region
175--300~MeV/$c^2$. The obtained bounds improve previous peak search results by two order of magnitude and the CERN PS191 results by order of magnitude in the selected heavy neutrino mass region.
In contrast to the CERN PS191 or BBN bounds the result is
model-independent because no assumptions about heavy neutrino
decay rates or couplings.

\begin{acknowledgments}

This research was supported in part by Grant \#14-12-00560 of the Russian
Science Foundation, the U.S. Department of Energy, the Ministry of Education,
Culture, Sports, Science and Technology of Japan through the Japan-U.S.
Cooperative Research Program in High Energy Physics and under 
Grant-in-Aids for
Scientific Research, the Natural Sciences and Engineering Research Council and
the National Research Council of Canada, National Natural Science Foundation of
China,
and the Tsinghua University Initiative Scientific Research Program.
\end{acknowledgments}

\appendix*
\section
{The photon veto acceptance measurement for the 
$K^+\to\mu^+\nu_\mu\gamma$ decay \label{subsec:spi}}

\begin{figure}[]
\centering
\includegraphics[width=\columnwidth]{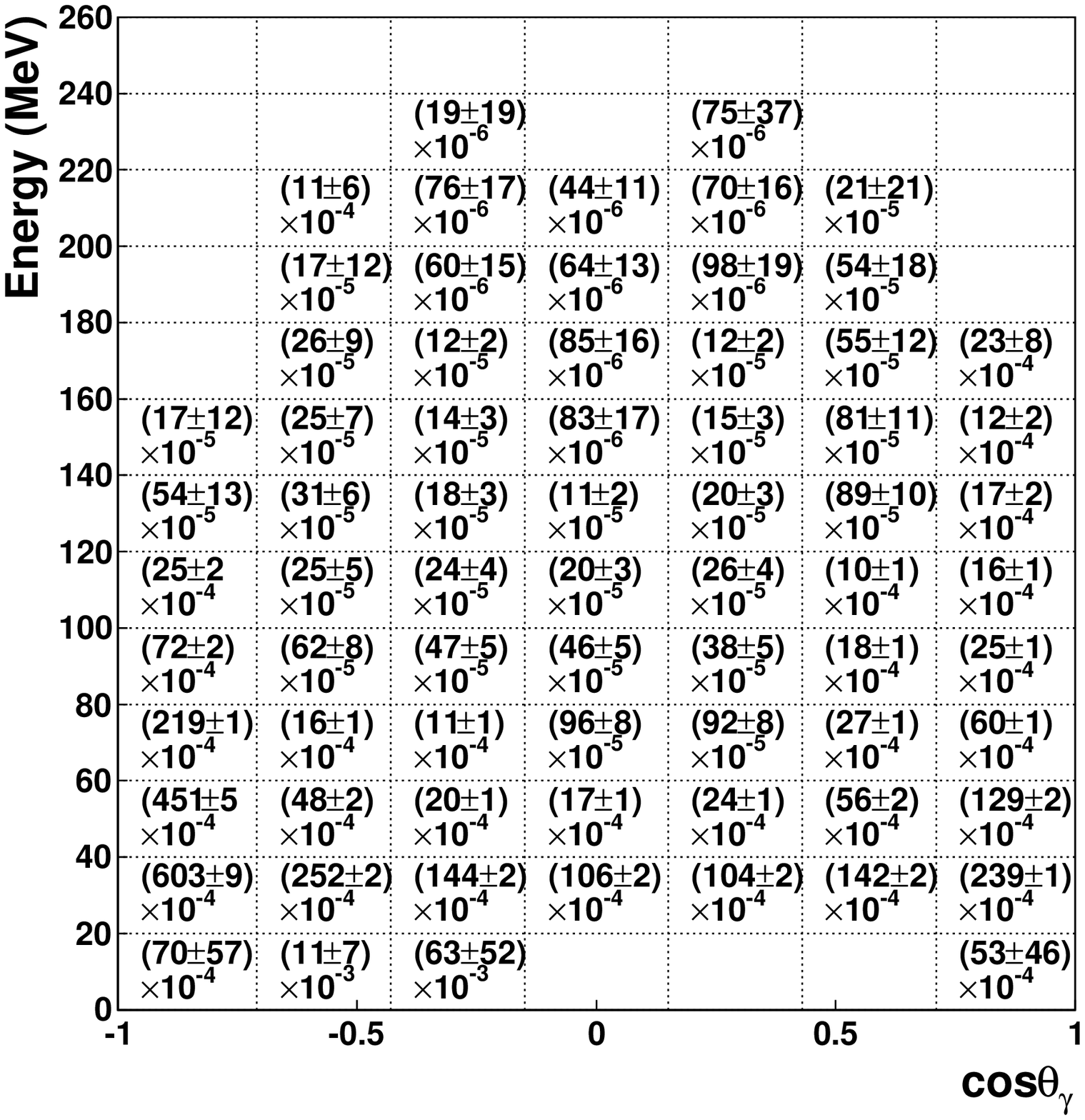}
\caption{Measured single photon inefficiency in the E949
detector~\cite{kentaro}. The angle between the outgoing
photon and beam directions~($z$-axis) is $\theta_\gamma$. The errors are statistical.}
\label{fig:spi1}
\end{figure}

The single
photon inefficiency table~(SPI) is shown in Figure~\ref{fig:spi1}.
The angle between the outgoing photon and beam
directions~($z$-axis) is $\theta_\gamma$.

To use this table we simulated the $\gamma$ direction and energy distributions
from $K^+\to\mu^+\nu_\mu\gamma$ decay.
The SPI table
includes both online and offline photon veto requirements and we cannot use it
if the muon and the photon hit the same stopping hextant because the range
stack photon veto was not applied in this case. A photon was rejected if it
hits any of the photon veto detectors with a detected energy more than 
1 MeV (offline PV threshold).
The true threshold for the photon energy should be lower than this because
of the contamination due to detector activity or electronic noise.
The photon acceptance was measured to be 
$A_{PV}=(1.24\pm 0.38)\times 10^{-2}$,
where the estimated uncertainty is determined by scanning the energy threshold
in the MC from 0 to 1 MeV.

\bibliography{e949_heavynu-first-round}

\end{document}